\documentclass[12pt]{article}
\usepackage[reqno]{amsmath}
\usepackage{bbm}
\usepackage{epsfig}
\usepackage{array}
\usepackage{float}
\usepackage{dsfont}
\usepackage{amstext}
\usepackage{amsfonts}

\usepackage{a4}

\usepackage{a4wide}
\def\be{\begin{equation}}
\def\ee{\end{equation}}
\def\gs{\mathrel{
   \rlap{\raise 0.511ex \hbox{$>$}}{\lower 0.511ex \hbox{$\sim$}}}}
\def\ls{\mathrel{
   \rlap{\raise 0.511ex \hbox{$<$}}{\lower 0.511ex \hbox{$\sim$}}}}

\newcommand{\onbb}{neutrinoless double beta decay }
\newcommand{\ba}{\begin{array}{c}}
\newcommand{\baz}{\begin{array}{cc}}
\newcommand{\bad}{\begin{array}{ccc}}
\newcommand{\bav}{\begin{array}{cccc}}
\newcommand{\bea}{\begin{equation} \begin{array}{c}}
\newcommand{\eea}{ \end{array} \end{equation}}
\newcommand{\ea}{\end{array}}

\newcommand{\dma}{\mbox{$\Delta m^2_{\rm A}$}}
\newcommand{\meff}{\mbox{$\langle m \rangle$}}

\newcommand{\sss}{\sin^2 \theta_{12}}
\newcommand{\sch}{\sin^2 \theta_{13}}

\begin{document}

\title{\vspace{-2cm}
\hfill {\small MPP--2006--140}\\
\vspace{-0.3cm} 
\hfill {\small hep--ph/0611030} 
\vskip 0.4cm
\bf 
On Symmetric Lepton Mixing Matrices
}
\author{
Kathrin A.~Hochmuth$^a$\thanks{email: \tt hochmuth@mppmu.mpg.de}~~~and~
Werner Rodejohann$^b$\thanks{email: \tt werner.rodejohann@mpi-hd.mpg.de} 
\\\\
{\normalsize \it $^a$Max--Planck--Institut f\"ur Physik
 (Werner--Heisenberg--Institut),}\\
{\normalsize \it F\"ohringer Ring 6, 
D--80805 M\"unchen, Germany}\\ \\ 
{\normalsize \it $^b$Max--Planck--Institut f\"ur Kernphysik,}\\
{\normalsize \it  Postfach 103980, D--69029 Heidelberg, Germany}
}
\date{}
\maketitle
\thispagestyle{empty}
\vspace{-0.8cm}
\begin{abstract}
\noindent  
Contrary to the quark mixing matrix, the lepton mixing matrix 
could be symmetric. We study the phenomenological consequences of this 
possibility. 
In particular, we find that symmetry would imply 
that $|U_{e3}|$ is larger than 0.16, i.e., above its current 2$\sigma$ limit. 
The other mixing angles are also constrained and 
$CP$ violating effects in neutrino oscillations are suppressed, even though 
$|U_{e3}|$ is sizable. Maximal atmospheric mixing is only allowed if the 
other observables are outside their current $3\sigma$ ranges, and  
$\sin^2 \theta_{23}$ lies typically below 0.5. 
The Majorana phases are not affected, but the implied values of 
the solar neutrino mixing angle have some effect on the predictions for 
neutrinoless double beta decay.
We further discuss some formal properties of a symmetric 
mixing matrix.

\end{abstract}

\newpage

Low energy neutrino physics \cite{reviews} is described by the 
neutrino mass matrix 
\be
m_\nu = U \, m_\nu^{\rm diag} \, U^T ~,
\ee 
where $U$ is the leptonic mixing, or 
Pontecorvo-Maki-Nakagawa-Sakata (PMNS) \cite{PMNS}, 
matrix in the basis in which the charged lepton mass 
matrix $m_\ell$ is real and diagonal. The three neutrino masses 
are contained in $m_\nu^{\rm diag} = {\rm diag}(m_1, m_2, m_3)$. 
A useful parameterization for the unitary PMNS matrix is  
\bea 
\label{eq:Upara}
U = 
\left( \bad 
U_{e1} & U_{e2} & U_{e3} \\[0.2cm] 
U_{\mu 1} & U_{\mu 2} & U_{\mu 3} \\[0.2cm] 
U_{\tau 1} & U_{\tau 2} & U_{\tau 3} 
\ea
\right) \\[0.3cm] 
= \left( \bad 
c_{12} \, c_{13} & s_{12} \, c_{13} & s_{13} \, e^{-i \delta}  \\[0.2cm] 
-s_{12} \, c_{23} - c_{12} \, s_{23} \, s_{13} \, e^{i \delta} 
& c_{12} \, c_{23} - s_{12} \, s_{23} \, s_{13} \, e^{i \delta} 
& s_{23} \, c_{13}  \\[0.2cm] 
s_{12} \, s_{23} - c_{12} \, c_{23} \, s_{13} \, e^{i \delta} & 
- c_{12} \, s_{23} - s_{12} \, c_{23} \, s_{13} \, e^{i \delta} 
& c_{23} \, c_{13}  \\ 
               \ea   
\right) 
 {\rm diag}(1, e^{i \alpha}, e^{i (\beta + \delta)}) \, , 
\eea 
where we have used the usual notations $c_{ij} = \cos\theta_{ij}$, 
$s_{ij} = \sin\theta_{ij}$ and introduced the Dirac $CP$-violating
phase $\delta$. There are two independent Majorana $CP$-violating 
phases $\alpha$ and $\beta$ \cite{BHP80} contained in the 
diagonal phase matrix 
$P = {\rm diag}(1, e^{i \alpha}, e^{i (\beta + \delta)})$. 
Various experiments and their analyzes  
revealed the following allowed 2, 3 and 4$\sigma$ 
ranges of the mixing angles \cite{thomas}:
\begin{eqnarray} \label{eq:data}
\sss &=& 0.30^{+0.06, \,0.10, \, 0.14}_{-0.04, \, 0.06,\, 0.08} 
~,\nonumber\\
\sin^2\theta_{23} &=& 0.50^{+0.13, \, 0.18,  \,0.21}
_{-0.12, \, 0.16,  \,0.19} ~,\\
\sch &<& 0.025~(0.041, \, 0.058)~.\nonumber
\end{eqnarray}
Note that the present best-fit value for $\sch$ is zero, 
and that there is no information on any of the phases.\\
 
Obviously, the precise form of the PMNS matrix will shed some light 
on the underlying theory of lepton flavor. One 
interesting possible property of the PMNS matrix is that it might 
be symmetric\footnote{Symmetry of the 
PMNS matrix around its $U_{e3}$--$U_{\mu 2}$--$U_{\tau 1}$--axis 
has been studied in \cite{sym}.}, $U = U^T$. 
In this letter we study in detail the consequences of this possibility. 
Recently, a symmetric PMNS matrix has been shown to follow 
from certain classes of models in which the very same unitary 
matrix is associated with the diagonalization of all 
fermion mass matrices \cite{QLU}. 
To conduct a more detailed phenomenological analysis of a symmetric PMNS matrix 
than the one performed in Ref.~\cite{QLU} is one of the 
motivations of this letter. However, to put the discussion on a broader basis, 
let us first comment on the formal properties of a symmetric PMNS matrix, 
which are similar to the properties of a symmetric 
CKM matrix\footnote{This possibility has been ruled out, see below.} 
\cite{branco,sarkar,CKMother}: 

\begin{itemize}
\item[(i)] first recall that in general $U = U_\ell^\dagger \, U_\nu$ holds, 
where $U_\ell$ is associated with the diagonalization of the 
charged lepton mass matrix via $m_\ell \, m_\ell^\dagger = 
U_\ell \, (m_\ell^{\rm diag})^2 \, U_\ell^\dagger$, 
with $m_\ell^{\rm diag} = (m_e, m_\mu, m_\tau)$. Hence, if 
\be 
U_\ell = S \, U_\nu^\ast~,
\ee
where $S$ is symmetric and unitary, then $U$ is also symmetric \cite{sarkar}. 
It holds that $S = U_\ell \, U_\nu^T$. 
If $S = \mathbbm{1}$ and $m_\ell$ is hermitian 
we recover the model from Ref.~\cite{QLU}. In this scenario we have   
$m_\nu = U_\nu \, m_\nu^{\rm diag} \, U_\nu^T$ 
and $m_\ell = U_\nu^\ast \, 
m_\ell^{\rm diag} \, U_\nu^T$. 
Another special case occurs if $S = \mathbbm{1}$ 
and $m_\ell$ is symmetric. 
Hereby we obtain 
$m_\nu^\ast = U_\nu^\ast \, m_\nu^{\rm diag} \, U_\nu^\dagger$ 
and $m_\ell = U_\nu^\ast \, 
m_\ell^{\rm diag} \, U_\nu^\dagger$, i.e., 
$m_\nu^\ast$ and $m_\ell$ are diagonalized by the same matrix;

\item[(ii)] another formal aspect is the following \cite{branco}: 
we can write any unitary matrix, in 
particular the PMNS matrix, as $U = X \, U^{\rm diag} \, X^\dagger$, 
where $X$ is unitary and $U^{\rm diag}$ contains the eigenvalues of 
$U$: $U^{\rm diag} = {\rm diag}(e^{i \phi_1}, e^{i \phi_2}, e^{i \phi_3})$. 
At first we assume non-degenerate eigenvalues. 
If $X$ is real, then $U$ is obviously symmetric. 
To turn the argument around, note that from 
$U = X \, U^{\rm diag} \, X^\dagger$ one can obtain  
$U \, X = X \,  U^{\rm diag}$ and -- in case of a symmetric $U$ -- that 
$U \, X^\ast = X^\ast \,  U^{\rm diag}$. 
Thus, the columns of $X$ and $X^\dagger$ are eigenvectors of $U$ with 
identical eigenvalues. Consequently, they only 
differ by a phase and we can write 
$X = O \, Q$, where $O$ is real and orthogonal and 
$Q = {\rm diag}(e^{i \omega_1},e^{i \omega_2}, e^{i \omega_3})$. 
However, from $U = X \, U^{\rm diag} \, X^\dagger$ it follows that 
multiplying $X$ with $Q^\dagger$ will lead to the same $U$ and hence 
the phases in $Q$ are unphysical. Thus, we have shown that a 
symmetric PMNS matrix with non-degenerate eigenvalues 
implies that its eigenvectors are real \cite{branco}. 
Suppose now that two of the eigenvalues of $U$ 
are degenerate: 
in this case, without loss of generality,  
$U^{\rm diag} = {\rm diag}(e^{i \phi_1}, e^{i \phi_1}, e^{i \phi_3})$, 
which can be written as 
$U^{\rm diag} = 
e^{i \phi_1} \, (\mathbbm{1} + {\rm diag}(0,0, e^{i (\phi_3 - \phi_1)} - 1))$. 
Simply evaluating $U = X \, U^{\rm diag} \, X^\dagger$ shows 
that $|U|$ is symmetric\footnote{The case of all three eigenvalues 
being identical is trivial.}. 
We will show next that, if $|U|$ is symmetric, rephasing of the lepton 
fields allows to make $U$ symmetric;

\item[(iii)]
it is a special feature of three fermion generations that a mixing matrix 
having symmetric moduli 
($|U_{e3}| = |U_{\tau 1}|$, $|U_{e2}| = |U_{\mu 1}|$ and 
$|U_{\mu3}| = |U_{\tau 2}|$) can be rephased in a way 
such that $\arg (U_{e3}) = \arg (U_{\tau 1})$, 
$\arg (U_{e2}) = \arg (U_{\mu 1})$ and 
$\arg (U_{\mu3}) = \arg (U_{\tau 2})$ \cite{branco}. To see this, consider a 
rephasing of the neutrino and charged lepton fields via 
$\nu_i \rightarrow \nu_i \, e^{i \sigma_i}$ and 
$\ell_j \rightarrow \ell_j \, e^{i \rho_j}$ with $i,j = e, \mu, \tau$ or 
$1, \, 2, \, 3$.  
As a consequence, the PMNS matrix element $U_{ij}$ is changed to 
$U_{ij} \, e^{i (\sigma_i - \rho_j)}$ and in addition the Majorana phases 
are modified:
${\rm diag}(1, e^{i \alpha}, e^{i (\beta + \delta)}) 
\rightarrow {\rm diag}(e^{i \sigma_1}, e^{i (\alpha + \sigma_2)}, 
e^{i (\beta + \delta + \sigma_3)})$. 
Suppose the arguments of $U_{ij}$ before rephasing are $\phi_{ij}$. In order 
to have $\arg (U_{ij}) = \arg (U_{ji})$ after rephasing, the 
parameters with which we rephase the lepton fields have to 
submit to the condition 
$\phi_{ij} - \phi_{ji} = \sigma_i - \sigma_j 
+ \rho_i - \rho_j \mod\!\!(2\pi)$. There is a solution for this condition 
if 
\be \label{eq:Im} 
{\rm Im} 
\{U_{e2} \, U_{e3}^\ast \, U_{\tau 2}^\ast \, U_{\mu 3} 
\, U_{\tau 1} \, U_{\mu 1}^\ast\} = 0~.
\ee 
It is trivial to show that this equation is automatically fulfilled due to 
unitarity of the PMNS matrix in the case of symmetric moduli: 
consider the unitarity relation 
for the second and third row and column of $U$: 
$U_{e2} \, U_{e3}^\ast + U_{\mu 2} \, U_{\mu 3}^\ast 
+ U_{\tau 2} \, U_{ \tau3}^\ast = 0$ and 
$U_{\mu 1} \, U_{\tau 1}^\ast + U_{\mu 2} \, U_{\tau 2}^\ast 
+ U_{\mu 3} \, U_{ \tau 3}^\ast = 0$. Multiplying the first expression with 
$U_{\tau 2}^\ast$ and the second with $U_{\mu 3}^\ast$ and subtracting the 
two resulting equations while assuming $|U_{\mu3}| = |U_{\tau 2}|$ yields 
$U_{e2} \, U_{e3}^\ast \, U_{\tau 2}^\ast = U_{\mu 1} \, 
\, U_{\tau 1}^\ast \, U_{\mu 3}^\ast$. This is again the 
condition in Eq.~(\ref{eq:Im}). One can show that for more than 
three fermion generations symmetric moduli are not automatically 
equivalent to a complete symmetric $U$ \cite{branco}. 
This would be of importance if 
the LSND result was confirmed; 

\item[(iv)]a related question is the number of constraints the assumption of 
a symmetric PMNS matrix imposes on the observables. Even though there are 
in principle three symmetry conditions,  
$|U_{e3}| = |U_{\tau 1}|$, $|U_{e2}| = |U_{\mu 1}|$ and 
$|U_{\mu3}| = |U_{\tau 2}|$, it is easy to see that as a consequence of 
unitarity
\be
|U_{e3}|^2 - |U_{\tau 1}|^2 = |U_{e2}|^2 - |U_{\mu 1}|^2 = 
|U_{\mu3}|^2 - |U_{\tau 2}|^2~ .
\ee
Therefore, only one constraint is inflicted on the neutrino 
mixing observables. 
\end{itemize}

Hence the message delivered by the last two pints is that 
in order to investigate the phenomenological consequences of a 
symmetric PMNS matrix it is 
obviously sufficient to consider symmetric moduli. 
This implies in particular 
that the Majorana phases are generally not subject to any constraint. 
Only the Dirac phase and the three mixing angles will be affected. 
Moreover, the same result for 
the observables will be obtained for all three symmetry conditions.\\

Note that a symmetric CKM matrix $V$ 
is ruled out with current data. 
For instance, one finds that \cite{ckm} 
$|V_{ub}| = (3.82^{+0.49}_{-0.44}) \cdot 10^{-3}$ 
yet $|V_{td}| = (8.28^{+1.38}_{-0.86}) \cdot 10^{-3}$, where the errors 
are at the $3\sigma$ level.  
In terms of the Wolfenstein parameterization \cite{wolf}, one has 
$|V_{ub}| = A \, \lambda^3 \, (\rho^2 + \eta^2)$ and 
$|V_{td}| = A \, \lambda^3 \, \left( (1 - \rho)^2 + \eta^2 \right)$. 
Since $\rho \neq \frac 12$ the two elements can not be equal.\\

From now on we will focus on the phenomenological consequences of 
a symmetric PMNS matrix. Though one could use a parameterization suitable 
for the study of symmetric matrices and identify the elements of  
this parameterization with the usual mixing angles and $CP$ phases  
from Eq.~(\ref{eq:Upara})\footnote{Such a parameterization will be  
useful if the PMNS matrix turns out to be indeed (close to) 
symmetric.}, we will 
focus in this letter on the usual parameterization of Eq.~(\ref{eq:Upara}) 
and directly obtain correlations between the neutrino mixing observables 
$\theta_{12} , \theta_{23} , \theta_{13}$ and $\delta$. 
First, let us obtain the ranges of the 
individual elements of the PMNS matrix: we vary the mixing angles in 
their allowed ranges given in Eq.~(\ref{eq:data}) 
and the phase $\delta$ between zero and $2\pi$, resulting in
\be \label{eq:ranges}
|U| = 
\left\{
\bad 
\left( 
\bad 
0.79 \div 0.86 & 0.50 \div 0.60 & 0 \div 0.16 \\
0.20 \div 0.55 & 0.40 \div 0.73 & 0.61 \div 0.80 \\
0.21 \div 0.56 & 0.42 \div 0.74 & 0.59 \div 0.79
\ea
\right) & \mbox{and } |J_{CP}| \le 0.037 & (\mbox{at }2\sigma)~,\\[0.3cm]
\left( 
\bad 
0.76 \div 0.87 & 0.48 \div 0.63 & 0 \div 0.20 \\
0.13 \div 0.60 & 0.33 \div 0.77 & 0.57 \div 0.82 \\
0.14 \div 0.61 & 0.35 \div 0.77 & 0.55 \div 0.81
\ea
\right) & \mbox{and } |J_{CP}| \le 0.048 & (\mbox{at }3\sigma)~,\\[0.3cm]
\left( 
\bad 
0.73 \div 0.88 & 0.46 \div 0.66 & 0 \div 0.24 \\
0.07 \div 0.65 & 0.27 \div 0.80 & 0.54 \div 0.84 \\
0.09 \div 0.66 & 0.28 \div 0.80 & 0.52 \div 0.83
\ea
\right) & \mbox{and } |J_{CP}| \le 0.056 &  (\mbox{at }4\sigma)~.
\ea
\right. 
\ee
We have also given the maximal possible value of the Jarlskog 
invariant $J_{CP}$ 
to which any $CP$ violating effect in neutrino oscillations  
is proportional \cite{JCP}:  
\be \label{eq:JCPlep} 
J_{CP} = {\rm Im} 
\left\{ U_{e1} \, U_{\mu 2} \, U_{e 2}^\ast \, U_{\mu 1}^\ast \right\} 
= \frac 18 \, \sin 2 \theta_{12} \, \sin 2 \theta_{23} \, 
\sin 2 \theta_{13} \, \cos \theta_{13} \, \sin \delta~. 
\ee
Note that the rephasing of the PMNS matrix elements as discussed 
before Eq.~(\ref{eq:Im}) leaves $J_{CP}$ invariant.\\

Consider now in Eq.~(\ref{eq:ranges}) the symmetry condition 
$|U_{e3}| = |U_{\tau 1}|$. Apparently, to fulfill this condition 
the current 2$\sigma$ ranges of the observables do not suffice.  
This can be easily understood qualitatively since $|U_{e3}|$ is given by 
$\sin \theta_{13}$ and is therefore small, whereas 
$|U_{\tau 1}| = |s_{12} \, s_{23} 
- c_{12} \, c_{23} \, s_{13} \, e^{i \delta}|$ is generally 
large. For a small $|U_{\tau 1}|$ of order $|U_{e3}|$ it 
is necessary that $\theta_{13}$ is large and that $\delta$ lies close to 
zero or $\pi$ in order to subtract the second term in $U_{\tau 1}$ 
from the first one. Moreover, $s_{12} \, s_{23} 
- c_{12} \, c_{23} \, s_{13}$ is smaller when $c_{23}$ is larger 
than $s_{23}$, i.e., atmospheric neutrino mixing will tend to be governed by 
$\sin^2 \theta_{23} < 1/2$. 
These statements can be made more precise: 
it is easy to proof that all three symmetry conditions, 
$|U_{e3}| = |U_{\tau 1}|$, $|U_{e2}| = |U_{\mu 1}|$ and 
$|U_{\mu3}| = |U_{\tau 2}|$ are fulfilled for one single 
condition, namely 
\be \label{eq:limit}
|U_{e3}| =  
\frac{\sin \theta_{12} \, \sin \theta_{23}}
{\sqrt{1 - \sin^2 \delta \, \cos^2 \theta_{12} \, \cos^2 \theta_{23}} 
+ \cos \delta \, \cos \theta_{12} \, \cos \theta_{23}}~.
\ee
Interestingly, this constraint can also be derived by setting the 
real and imaginary parts of $U_{e3} = U_{\tau 1}$ equal. 
One gets then two relations, 
$s_{13} = s_{12} \, s_{23}/(c_\beta + c_\delta \, c_{12} \, c_{23})$
and 
$(s_\beta + c_{12} \, c_{23} \, s_\delta) \, s_{13} = 0$. 
From the first one it follows that vanishing of $s_{13}$ in not possible, 
since experiments show that $s_{12}$ and $s_{23}$ cannot be zero. 
Realizing this, the second 
relation is fulfilled when 
$\sin \beta = -\sin \delta \, \cos \theta_{12} \, \cos \theta_{23}$. 
Inserting this expression in the first condition yields 
Eq.~(\ref{eq:limit}). However, care has to be taken when working with 
real and imaginary parts of mixing matrix elements, since their individual 
phases have no physical meaning.\\

With the 2 (3 and 4)$\sigma$ ranges of $\theta_{12}$ and $\theta_{23}$, 
and with varying the phase $\delta$ between zero and $2\pi$, 
one obtains from Eq.~(\ref{eq:limit}) 
that $|U_{e3}|^2 \gs 0.035$ (0.028 and 0.023), which has to be compared to 
the experimental upper limits of 0.025 (0.041 and 0.058). 
Therefore, a symmetric 
PMNS matrix predicts that $|U_{e3}|$ should be above its current 
2$\sigma$ limit. Consequently, the scenario is easily falsifiable, 
since the indicated value of $|U_{e3}|$ should be verified in upcoming 
measurements, in particular by the  
Double Chooz experiment \cite{newexp} (see also \cite{chef}): 
according to Ref.~\cite{newexp}, data taking can 
start in 2008 and the $3\sigma$ limit on 
$|U_{e3}|$ will be improved 
from its current value 0.04 to 0.01~(0.006) after 2~(6) years of data taking. 
These numbers are well below the prediction of a symmetric PMNS matrix. 
Fixing $\theta_{23}$ to $\pi/4$, gives 
$|U_{e3}|^2 \gs 0.050$ (0.046 and 0.042). 
This means that maximal atmospheric neutrino mixing would be 
at roughly 3 standard deviations in conflict with a symmetric PMNS matrix. 
Fixing $\theta_{12}$ 
to its best-fit point gives 
$|U_{e3}|^2 \gs 0.041$ (0.036 and 0.033), which is hardly compatible with the 
current 3$\sigma$ limit of $|U_{e3}|^2$. Moreover, maximal atmospheric mixing 
is not compatible with $\sin^2 \theta_{12} = 0.30$. 
We conclude that the current best-fit values of the oscillation parameters are 
not compatible with a symmetric PMNS matrix. To be precise, the prediction 
of $\sin^2 \theta_{12}= 0.30$ and $\sin^2 \theta_{23}= 1/2$ would be 
$|U_{e3}|^2 \gs 0.059$. 
To quantify the compatibility of a symmetric mixing matrix with current data 
we have also performed a simple $\chi^2$ minimization: we introduce  
\[ 
\chi^2 = \displaystyle\sum_{ij=12, 13, 23} 
\frac{ \left(s_{ij}^2 - (s_{ij}^2)_{\mbox{\scriptsize best-fit}} \right)^2}
{\sigma_{ij}^2}~,
\] 
where $(s_{ij}^2)_{\mbox{\scriptsize best-fit}}$ and 
$\sigma_{ij}$ are the best-fit values 
and errors from Eq.~(\ref{eq:data}). Obeying the symmetry condition 
Eq.~(\ref{eq:limit}), one can find a minimum of $\chi^2 = 10.29$ 
for the parameters 
$\sin^2 \theta_{12} \simeq 0.28$, $\sin^2 \theta_{23} \simeq 0.36$, 
$|U_{e3}|^2 \simeq 0.035$ and $\delta \simeq 0$. The corresponding pulls 
for $\sin^2 \theta_{12}$, $\sin^2 \theta_{23}$, and $|U_{e3}|^2$ are 
$-0.81$, $-1.78$ and $2.55$, respectively. \\

\begin{figure}[htb]\vspace{-0.51cm}
\begin{tabular}{cc}
\epsfig{file=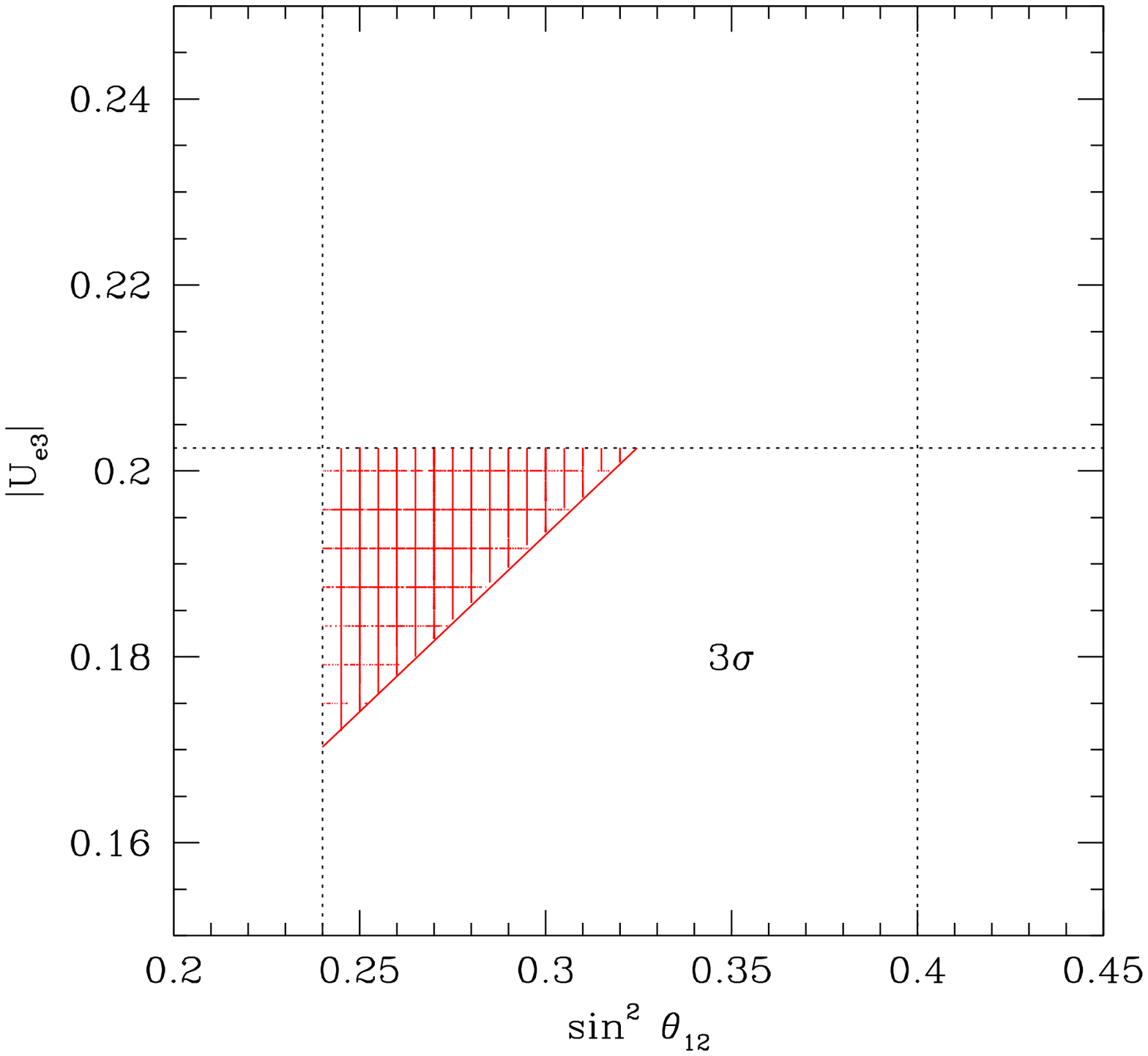,width=8cm,height=7cm} & 
\epsfig{file=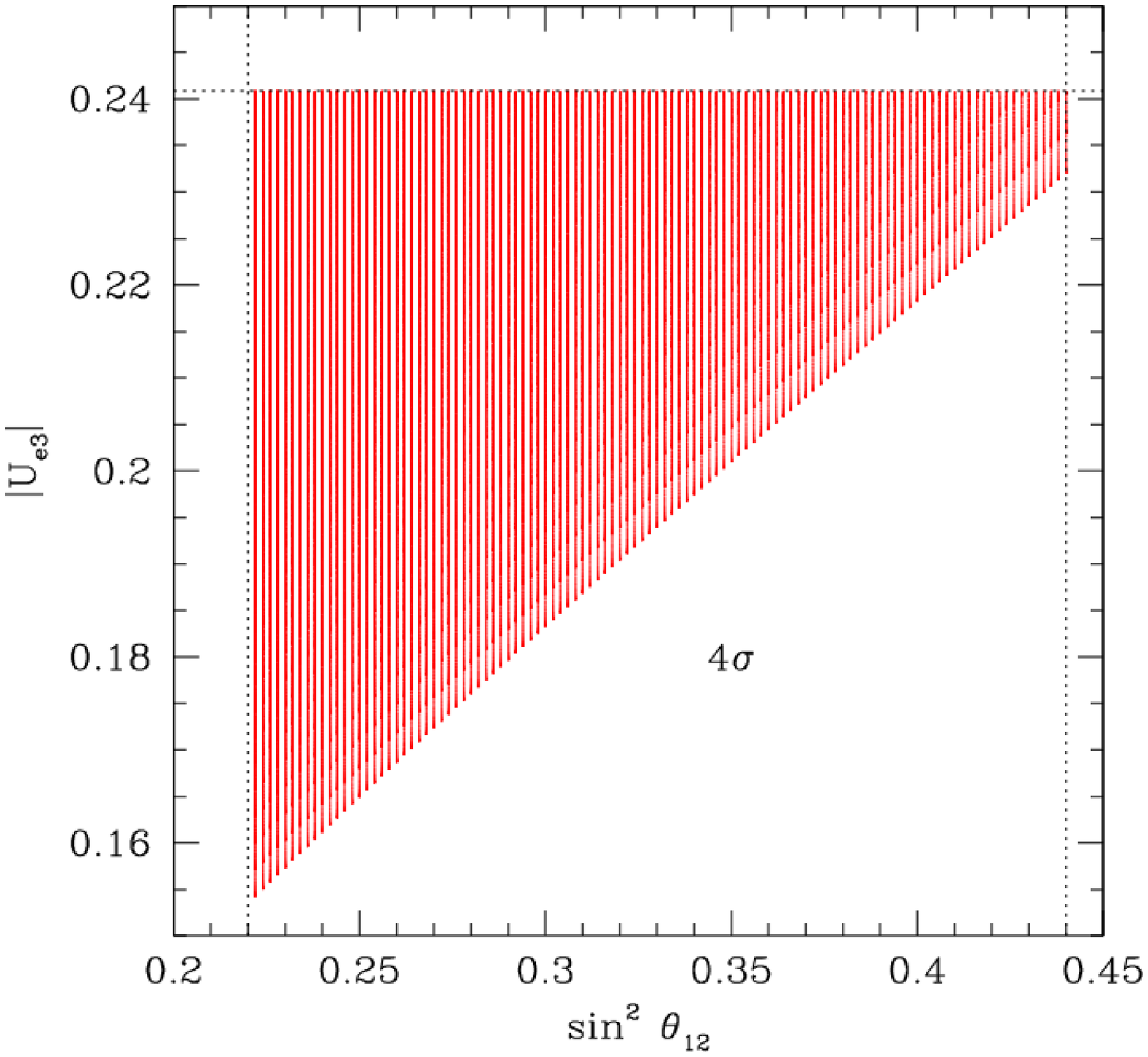,width=8cm,height=7cm} \\ 
\epsfig{file=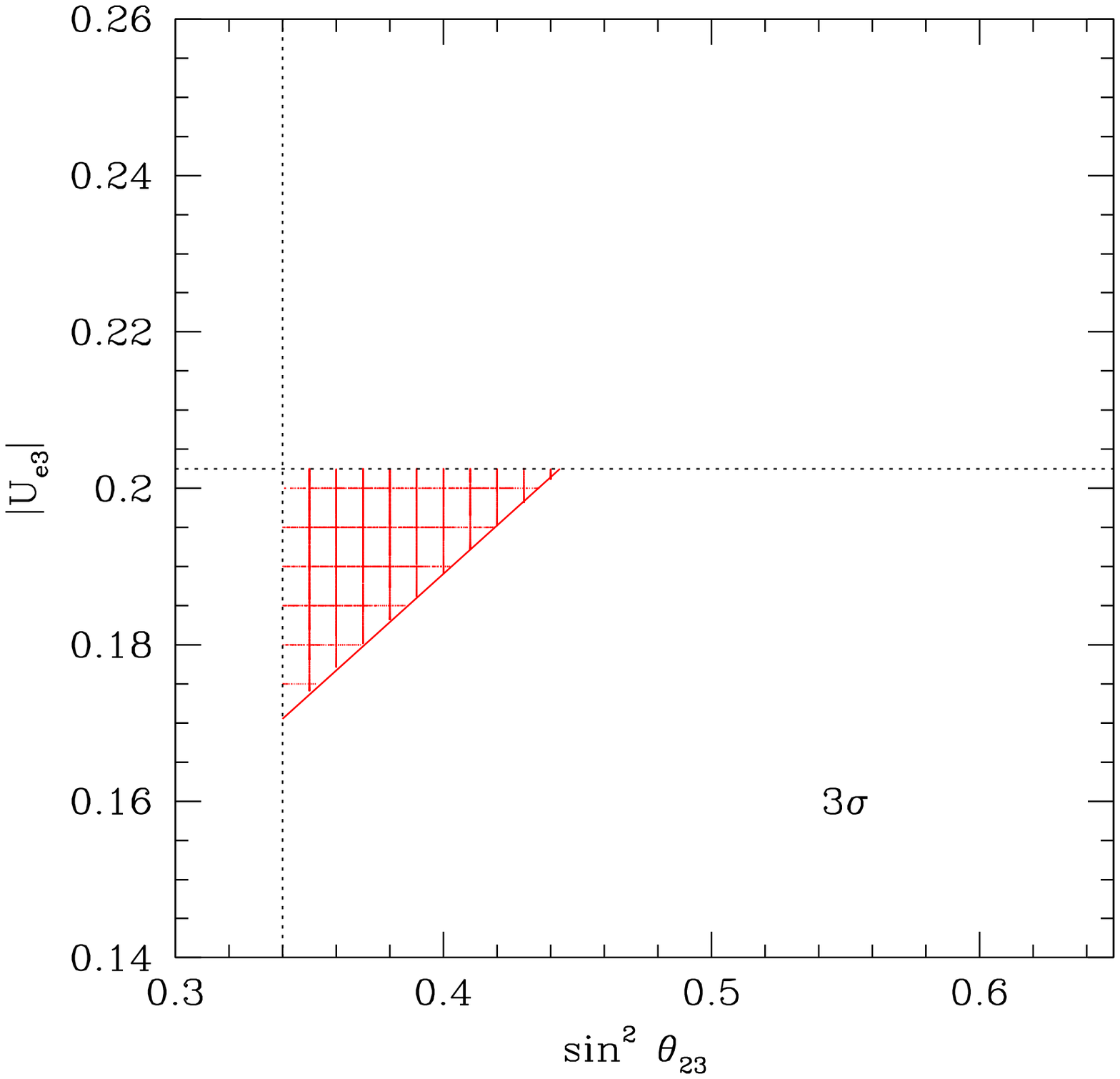,width=8cm,height=7cm} & 
\epsfig{file=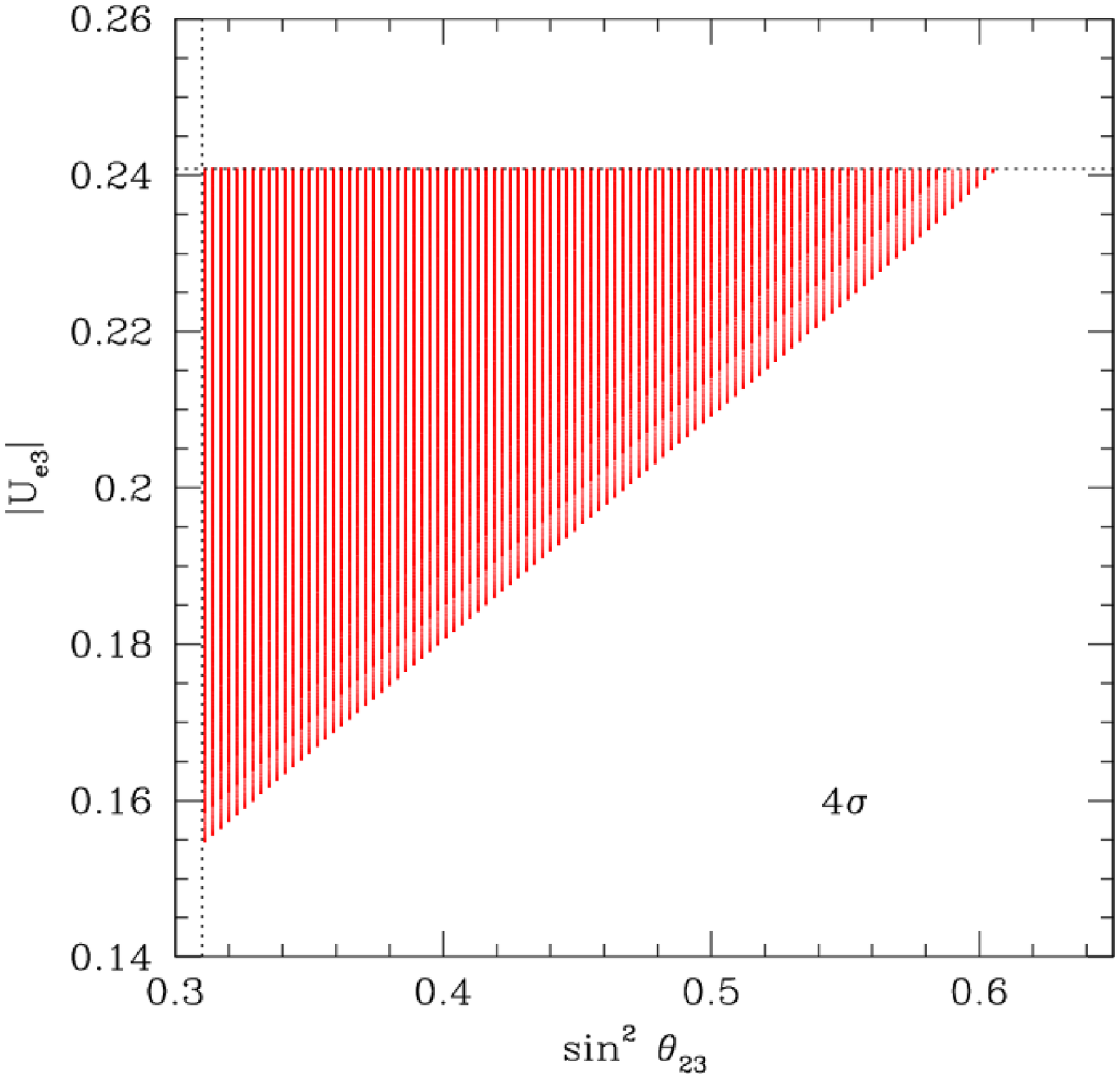,width=8cm,height=7cm} \\
\epsfig{file=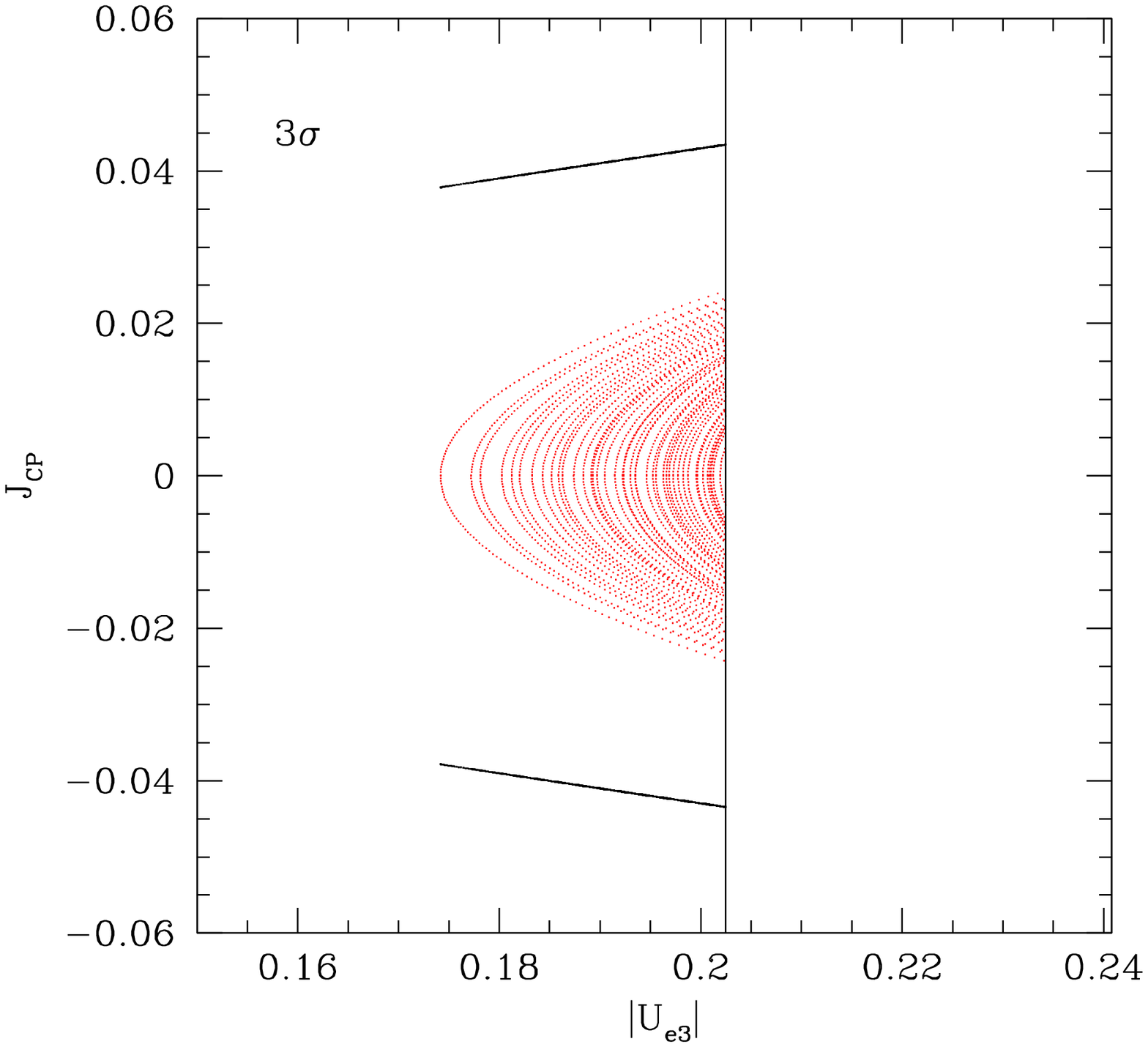,width=8cm,height=7cm} & 
\epsfig{file=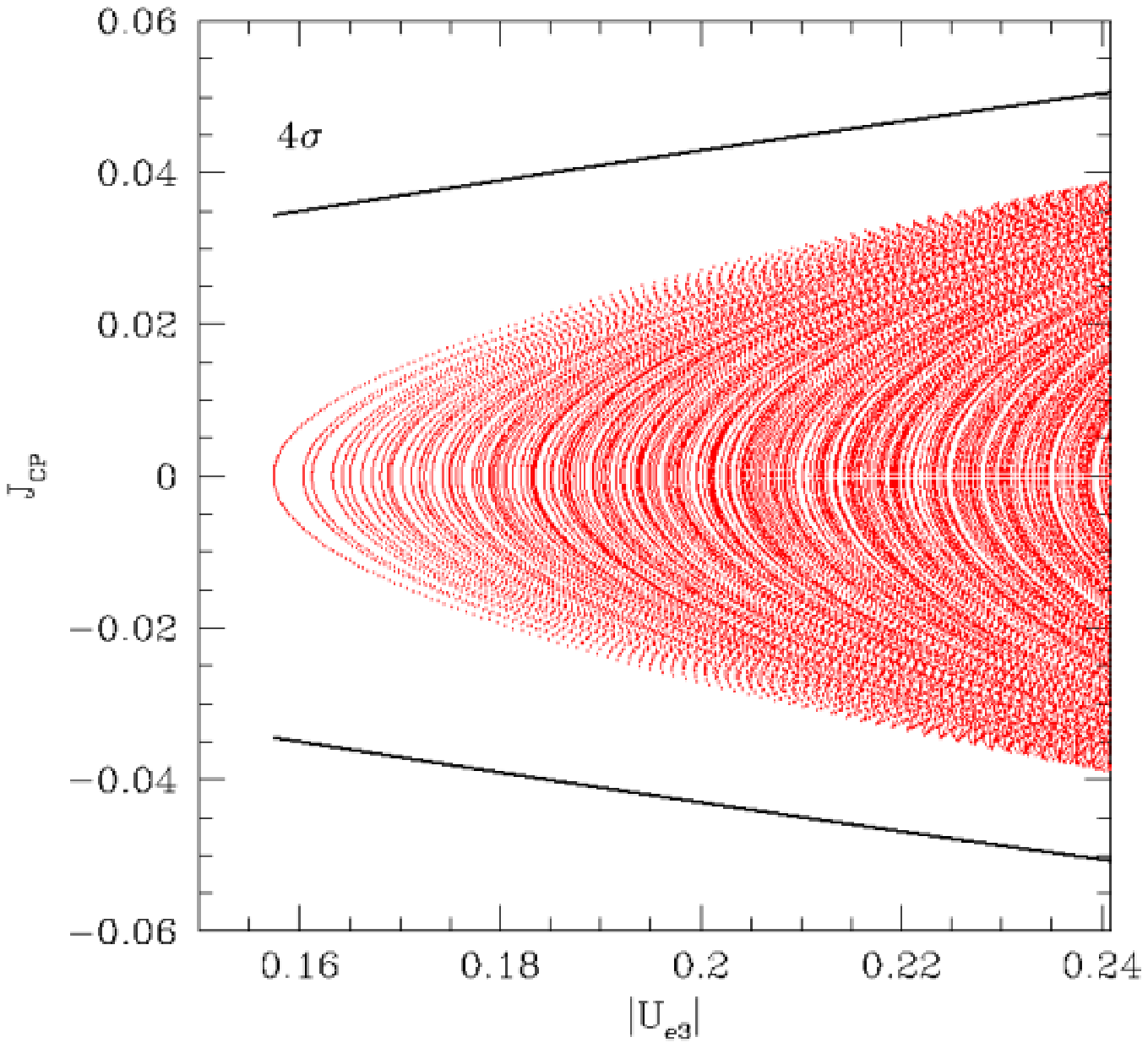,width=8cm,height=7cm}
\end{tabular}
\caption{\label{fig:osc}Plots of the oscillation observables 
for a symmetric PMNS matrix. The correlations are a 
consequence of Eq.~(\ref{eq:limit}). 
We allowed the parameters to vary in their current 
3 (left plots) and 4$\sigma$ (right plots) ranges, which are 
indicated in the plot.}
\end{figure}

In Fig.~\ref{fig:osc} we show plots of $\sin^2 \theta_{12}$, 
$\sin^2 \theta_{23}$ and $|J_{CP}|$ against $|U_{e3}|$, obtained from 
Eq.~(\ref{eq:limit}). When we simply vary 
the mixing angles $\theta_{12,13,23}$ in their allowed 
3 and 4$\sigma$ ranges from Eq.~(\ref{eq:data}) and require the PMNS 
matrix to be symmetric, the plots look identical. 
Note that the $3\sigma$ ranges of the oscillation parameters 
imply $\sin^2 \theta_{23} < 1/2$.  
In the plot of $|U_{e3}|$ against $J_{CP}$, we also indicated the 
maximal $|J_{CP}|$ allowed by current data. This serves to illustrate  
that $CP$ violating effects are rather small when the 
PMNS matrix is symmetric, even though $|U_{e3}|$ is sizable. 
Indeed, the scenario is compatible (at $3\sigma$) 
with $CP$ conservation, in which case $|U_{e3}|^2 \gs 0.035$ (0.028 and 0.023). 
Note further that, for the $3\sigma$ ranges of the oscillation 
parameters,    
$\sin^2 \theta_{12}$ takes values on the lower side of its 
allowed range. This has interesting implications for the effective 
mass $\meff = |\sum U_{ei}^2 \, m_i|$ 
governing \onbb$\!\!$. If neutrinos enjoy an inverted ordering, 
the minimal value of the effective mass is 
$\meff_{\rm IH}^{\rm min} = 
c_{13}^2 \, \sqrt{\dma} \, \cos 2 \theta_{12}$. Hence, the 
smaller $\theta_{12}$, the larger the minimal value of \meff~in the inverted 
ordering. This in turn simplifies distinguishing the normal from 
the inverted hierarchy with \onbb$\!\!$, or fully 
probing the inverted ordering regime \cite{0vbb}. 
To quantify this statement, 
the lower limit in case of an inverted hierarchy is in general 
$\meff_{\rm IH}^{\rm min}  \simeq 0.2 \, c_{13}^2 \, \sqrt{\dma}$, 
where we have inserted the lowest possible value of 
$\cos 2 \theta_{12}$ at $3\sigma$. 
With the constraint stemming from a symmetric PMNS matrix we see from 
Fig.~\ref{fig:osc} that $\sin^2 \theta_{12} \ls 0.32$ and consequently 
$\meff_{\rm IH}^{\rm min} \simeq 0.36 \, c_{13}^2 \, \sqrt{\dma}$. 
This is almost a factor of two larger than without the constraint 
from symmetry.\\


To summarize, we studied what consequences would arise from a 
symmetric PMNS matrix $U$, a scenario which in contrast to a 
symmetric CKM matrix 
is still compatible with data. 
We noted that in this case either the eigenvectors of $U$ are 
real or $U$ has two degenerate eigenvalues. 
A symmetric $U$ arises when $U_\ell$ and $U_\nu$ are 
connected by a symmetric and unitary 
matrix $S$ via $U_\ell = S \, U_\nu^\ast$. One simple example is when 
the neutrino mass matrix and the complex conjugate of a symmetric 
charged lepton mass matrix are diagonalized by the same matrix. 
Symmetry implies one constraint on the neutrino oscillation 
observables, not on the Majorana phases. The scenario is easily falsifiable 
in the near future, since it predicts that $|U_{e3}|$ 
is larger than its current 
2$\sigma$ limit of about 0.16. Experiments like Double Chooz 
can therefore easily rule out a symmetric PMNS matrix. 
In addition, there are interesting and 
testable correlations between the observables, as given in 
Eq.~(\ref{eq:limit}) and illustrated in 
Fig.~\ref{fig:osc}. If the 3$\sigma$ ranges of the oscillation parameters 
are taken, $\theta_{23}$ 
cannot be maximal and lies below $\pi/4$. 
Solar neutrino mixing is far away from its maximal 
allowed value, which affects predictions for \onbb$\!\!$. 
In general, the $CP$ phase $\delta$ is close 
to a $CP$ conserving value. 


\vspace{0.5cm}
\begin{center}
{\bf Acknowledgments}
\end{center}
We thank G.~Branco for encouragement. 
This work was supported by the ``Deutsche Forschungsgemeinschaft'' 
in the ``Sonderforschungsbereich 375 f\"ur Astroteilchenphysik'' 
(K.A.H) and under project number RO--2516/3--1 (W.R.).

\pagestyle{empty}

\begin{thebibliography}{99} 

\bibitem{reviews}
For recent reviews, see 
R.~N.~Mohapatra {\it et al.},
hep-ph/0510213; 
R.~N.~Mohapatra and A.~Y.~Smirnov,
hep-ph/0603118; 
A.~Strumia and F.~Vissani,
  hep-ph/0606054.


\bibitem{PMNS} B. Pontecorvo, 
Sov.\ Phys.\ JETP {\bf 6}, 429 (1957)
  [Zh.\ Eksp.\ Teor.\ Fiz.\  {\bf 33}, 549 (1957)];
Sov.\ Phys.\ JETP {\bf 7}, 172 (1958)
  [Zh.\ Eksp.\ Teor.\ Fiz.\  {\bf 34}, 247 (1957)]; 
Z. Maki, M. Nakagawa and S. Sakata,
Prog.\ Theor.\ Phys.\  {\bf 28}, 870 (1962).

\bibitem{BHP80}S.~M.~Bilenky, J.~Hosek and S.~T.~Petcov,
  Phys.\ Lett.\ B {\bf 94}, 495 (1980); 
J.~Schechter and J.~W.~F.~Valle,
  Phys.\ Rev.\ D {\bf 22}, 2227 (1980);  
  Phys.\ Rev.\ D {\bf 23}, 1666 (1981); 
  M.~Doi {\it et al.},
  Phys.\ Lett.\ B {\bf 102}, 323 (1981). 


\bibitem{thomas}T.~Schwetz, 
  hep-ph/0606060; 
  M.~Maltoni {\it et al.}, 
  hep-ph/0405172v5. 



\bibitem{sym}Z.~Z.~Xing,
  Phys.\ Rev.\ D {\bf 65}, 113010 (2002). 


\bibitem{QLU}A.~S.~Joshipura and A.~Y.~Smirnov,
 Nucl.\ Phys.\ B {\bf 750}, 28 (2006).

\bibitem{branco}G.~C.~Branco and P.~A.~Parada,
  Phys.\ Rev.\ D {\bf 44}, 923 (1991). 

\bibitem{sarkar}M.~K.~Samal and U.~Sarkar,
  Phys.\ Rev.\ D {\bf 45}, 2421 (1992).



\bibitem{CKMother}Other aspects of symmetric CKM matrices were 
studied, e.g., in 
P.~Kielanowski,
Phys.\ Rev.\ Lett.\  {\bf 63}, 2189 (1989); 
J.~L.~Rosner,
  Phys.\ Rev.\ Lett.\  {\bf 64}, 2590 (1990);
 M.~K.~Samal, D.~Choudhury, U.~Sarkar and R.~B.~Mann,
  Phys.\ Rev.\ D {\bf 44}, 2860 (1991);
M.~Tanimoto,
  Mod.\ Phys.\ Lett.\ A {\bf 8}, 1387 (1993); 
H.~Gonzalez, S.~R.~Juarez W., P.~Kielanowski and G.~Lopez Castro,
  Phys.\ Lett.\ B {\bf 440}, 94 (1998); 
Z.~Z.~Xing,
  J.\ Phys.\ G {\bf 23}, 717 (1997);
S.~Chaturvedi and V.~Gupta,
  Mod.\ Phys.\ Lett.\ A {\bf 19}, 159 (2004). 

\bibitem{ckm}
J.~Charles {\it et al.}  [CKMfitter Group],
  Eur.\ Phys.\ J.\ C {\bf 41}, 1 (2005); 
A.~Hocker and Z.~Ligeti,
  hep-ph/0605217.

\bibitem{wolf}L.~Wolfenstein,
  Phys.\ Rev.\ Lett.\  {\bf 51}, 1945 (1983).


\bibitem{JCP}C.~Jarlskog, 
Phys.\ Rev.\ Lett.\  {\bf 55}, 1039 (1985); 
see also P.~I.~Krastev and S.~T.~Petcov,
  Phys.\ Lett.\ B {\bf 205}, 84 (1988).

\bibitem{newexp}
F.~Ardellier {\it et al.}  [Double Chooz Collaboration],
  hep-ex/0606025. 

\bibitem{chef}
  P.~Huber, M.~Lindner, M.~Rolinec, T.~Schwetz and W.~Winter,
  Phys.\ Rev.\ D {\bf 70}, 073014 (2004). 


\bibitem{0vbb}Recent analyzes of distinguishing neutrino mass 
hierarchies with \onbb are: 
S.~Pascoli, S.~T.~Petcov and T.~Schwetz,
  Nucl.\ Phys.\ B {\bf 734}, 24 (2006); 
S.~Choubey and W.~Rodejohann,
  Phys.\ Rev.\ D {\bf 72}, 033016 (2005); 
A.~de Gouvea and J.~Jenkins,
  hep-ph/0507021; 
M.~Lindner, A.~Merle and W.~Rodejohann,
  Phys.\ Rev.\ D {\bf 73}, 053005 (2006). 

\end{thebibliography}
\end{document}